\documentclass{article}
\usepackage{e,epsfig}
\begin{document}
\branch{A}
\DOI{123}                       
\idline{A}{1, 1--11}{1}         
\editorial{}{}{}{}              
\newcommand{\pom}{{\rm I\! P}}
\newcommand{\reg}{{\rm I\! R}}
\newcommand{\lsim}{\raisebox{-.6ex}{${\textstyle\stackrel{<}{\sim}}$}}
\newcommand{\gsim}{\raisebox{-.6ex}{${\textstyle\stackrel{>}{\sim}}$}}
\newcommand{\ptmiss}{\mbox{$\not\hspace{-1mm}\pt$}} 
\newcommand{\ptmisso}{\mbox{$\not\hspace{-1mm}\pt^{~outer}$}} 
\newcommand{\ecalP}{\mbox{$\frac{Q^2}{Q^2+M_Z^2}$}} 
\newcommand{\MWtwo}{\mbox{$M^2_{\tiny{W}}$}} 
\newcommand{\MZtwo}{\mbox{$M^2_{\tiny{Z}}$}} 
\newcommand{\MZ}{\mbox{$M_{\tiny{Z}}$}} 
\newcommand{\eplus}{\mbox{$\rm e^+$}} 
\newcommand{\emin}{\mbox{$\rm e^-$}} 
\newcommand{\ee}{\eplus \emin} 
\newcommand{\Jpsi}{\mbox{$J/ \psi$}} 
\def\coll{Collaboration}
\def\etal{et al.}
\title{Visions of $ep$ physics}
\author{B. Foster\inst{1,2}}
\institute{H.H. Wills Physics Laboratory, University of Bristol, Tyndall Avenue,
Bristol, BS8 1TL, U.K. \\
e-mail: b.foster@bris.ac.uk
\and
DESY, Notkestrasse 85, 22607 Hamburg, Germany.}
%
%
\maketitle
\begin{abstract}
The subject of lepton-hadron  scattering is discussed from its earliest
beginnings, concentrating on what we have learnt from the HERA electron-proton
storage ring. A brief selection of the HERA I results most relevant to LHC
are discussed. The HERA and ZEUS upgrades are outlined, together with the 
HERA II physics programme. The impact of HERA results on LHC is discussed, in
particular in the areas of luminosity measurement, background estimates and
possible signatures of new physics. Finally, possible future developments in 
lepton-proton physics beyond HERA II are discussed. 

\end{abstract}

%
\section{Introduction}
\label{sec-int}

The foundations of the Standard Model consist of a small number of deep
theoretical insights based on key experimental observations. Undoubtedly, the
quark-parton model and the deep inelastic scattering (DIS) experiments
begun at SLAC in the late 1960s are part of these key foundations. Of course,
the scattering of energetic ``simple'' particles from an unknown target
to elucidate its structure is an experimental technique with a long and distinguished history. One of the earliest, and perhpa sthe most famous, is the scattering of alpha particles from a thin gold foil carried out by Geiger and 
Marsden in Manchester in 1909, which led to the concept of the nuclear 
atom~\cite{pm:21:669,prslon:82:495}.

In this talk, I will briefly summarise the current status of $ep$
physics, concentrating on our knowledge of the proton structure, which, over
the majority of the currently explored phase space, comes primarily from
HERA results. I will then outline the HERA-II programme and how
the ZEUS and H1 experiments have been modified to take advantage of the large increase in luminosity. The main physics aims of the HERA-II programme
are outlined. I then discuss the importance and
relevance of the HERA-I and HERA-II output to
the physics programme of LHC. I conclude with a summary of possible future
facilities for $ep$ physics and a perspective on the future of
this field. 

\section{Current status of $ep$ physics and HERA-I results}
\label{sec-HERAI}
\begin{figure}[h]
\begin{center}
\epsfig{file=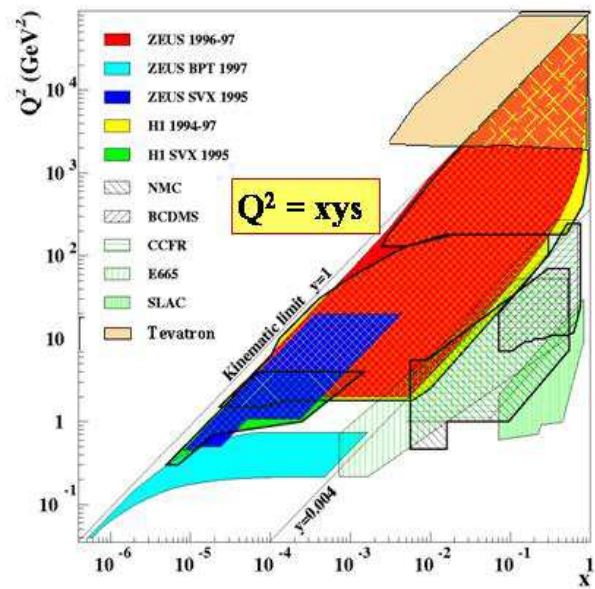,%
      width=8cm,%
      height=8cm%
        }
\end{center}
\caption{The kinematic plane in $x$ and $Q^2$ for experiments probing
the parton distribution of the proton. The regions explored by each
experiment are shown in a variety of shadings as shown in the legend.
Hadron-hadron collisions are also able to measure the proton structure,
predominantly at high $x$ and high $Q^2$.}
\label{fig:q2xkinreg}
\end{figure}

With the advent of the HERA electron-proton collider, the
explorable phase space in the kinematic invariants $Q^2$ (the virtuality of the
exchanged virtual photon) and $x$ (the fractional momentum of the parton
involved in the scattering has increased by approximately three orders of magnitude in each variable compared to what was available at earlier 
fixed-target experiments (see figure~\ref{fig:q2xkinreg}).
This extension in kinematic range has opened up qualitatively new fields of
study, both at high and low $Q^2$.

\vspace{1.5cm} 

\subsection{The proton structure function}
\label{sec-structurefn}

\begin{figure}[h]
\begin{center}
\epsfig{file=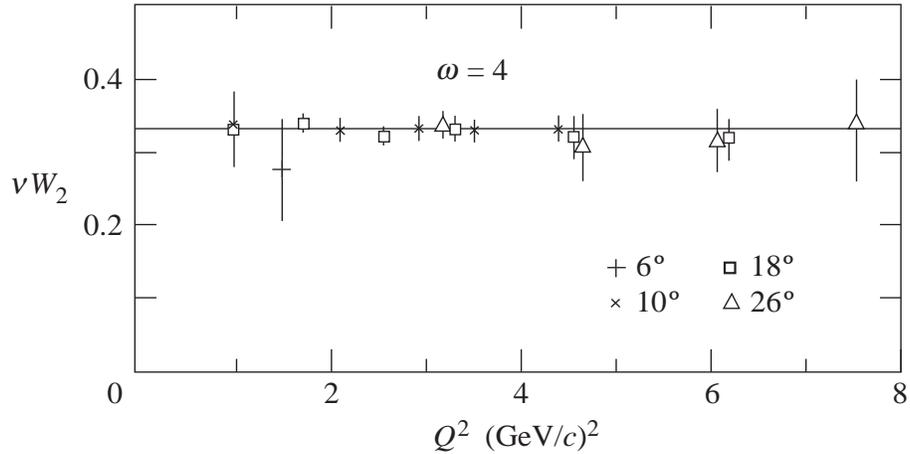,%
     width=12cm,%
      height=6cm%
        }
\end{center}
\caption{The $\nu W_2$ structure function at $\omega = 1/x = 4$ as a function
of $Q^2$ as measured by the SLAC-MITgroup\cite{SLACscaling}. Data taken at
four different scattering angles are shown. All data is consistent with
being independent of $Q^2$.}
\label{fig:SLACscale}
\end{figure}

Undoubtedly the most important observation of the SLAC DIS experiments
was that of scaling, as illustrated in figure~\ref{fig:SLACscale}. 
This shows the
$\nu W_2$ structure function (nowadays known as $F_2$) at 
$\omega (=x^{-1}) = 4$ 
as a function of $Q^2$. The remarkable lack of variation of $\nu W_2$ with
$Q^2$ is apparent; it was this that led directly to the postulation by
Feynman of the parton model. The interpretation of
these data, in retrospect, seems straight-forward. Since $Q^2$ is 
approximately proportional to the scattering angle of the electron and the structure function is proportional
to the scattering cross section, the flatness corresponds to the production
of an excess of particles at large angle compared to that expected from the
rapid fall off from the form factor of an extended object. The solution is
exactly the same as that which presented itself to Rutherford - the incident particle 
scatters not from an extended ``fuzzy'' object, the atom, but from
point-like scattering centres situated inside it. In his case this was
the nucleus; in the case of the SLAC experiment, the point-like scatterers
were the quarks.

\begin{figure}[h]
\begin{center}
\epsfig{file=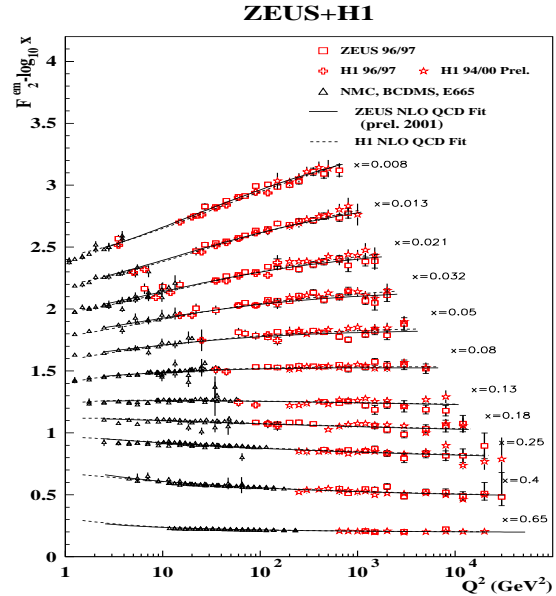,%
      width=8cm,%
      height=8cm%
        }
\end{center}
\caption{The $F_2$ structure function as measured by the H1 and ZEUS
experiments for bins at high $x$ as a function of $Q^2$. The bins centred around $x=0.25$ are where scaling
was originally observed in the SLAC experiments. Clear scaling violation
is observed in the HERA data outside this region,
particularly at low $x$.}
\label{fig:HERA-F2-xbins}
\end{figure}

That this scaling is still a feature of today's deep inelastic experiments
is illustrated by figure~\ref{fig:HERA-F2-xbins}. In modern terminology, the
deep inelastic double-differential neutral current cross section is expressed in terms of structure functions as 
\begin{eqnarray}  
\frac{d^2 \sigma} 
{dx dQ^2} & = & \frac{2\pi \alpha^2}{x Q^4}   
 \left[   
Y_+ \cdot
 F_2(x, Q^2) \right. \nonumber \\
 & \; - & \left.  {y^2} F_L(x, Q^2) \pm Y_- \cdot x F_3(x, Q^2) 
\right], 
\label{eq:Fl:sigma} 
\end{eqnarray}
where the $\pm$ before $xF_3$
is taken as positive for electron scattering and negative for 
positron scattering and $Y_{\pm}$ are kinematic factors given by 
\begin{equation}
Y_{\pm} = 1 \pm (1-y)^2,
\label{eq:Y}
\end{equation}
where $y$ is the inelasticity of the interaction. 
This can be expressed in terms of the other invariants as
\begin{eqnarray}
y &=& \frac{Q^2}{sx}
\label{eq:yq2sx} 
\end{eqnarray}
where $s$ is  the squared centre-of-mass energy of the $ep$ system.
The $F_2$ structure function can be expressed, in the ``DIS scheme'' in
a particularly simple way as
\begin{eqnarray} 
 F_2(x, Q^2) & = & \sum_{i=u,d,s,c,b} A_i(Q^2) \left[ xq_i(x,Q^2) + 
   x\overline{q}_i(x,Q^2) \right]  
\label{eq:F2:qpm}  
\end{eqnarray} 
The parton distributions $q_i(x,Q^2)$ and $\overline{q}_i(x,Q^2)$ refer  
to quarks and antiquarks of type 
$i$. For $Q^2 \ll M_Z^2$, where $M_Z$ is the mass of
the $Z^0$ boson, the quantities
$A_i(Q^2)$ are given by the square of the electric charge of quark or 
antiquark 
$i$. 

\begin{figure}[h]
\vskip1.5cm
\begin{center}
\epsfig{file=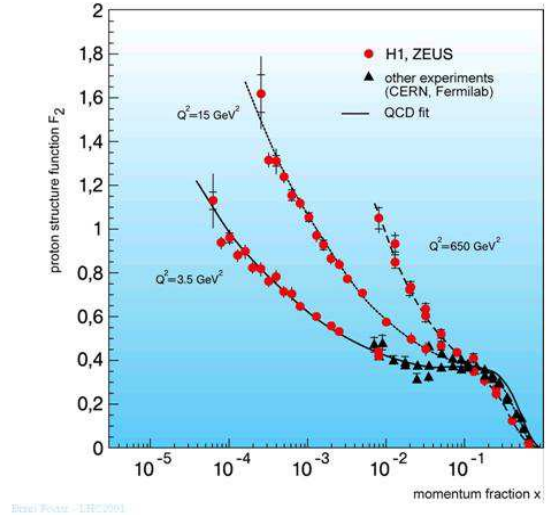,%
      width=7cm,%
      height=7cm%
        }
\end{center}
\caption{H1 and ZEUS data on the $F_2$ structure function shown in three
bins of $Q^2$ as a function of $x$. The steep rise of the structure
function at low $x$ is clearly apparent. The HERA data are now
as accurate as the fixed-target data and match onto it well.}
\label{fig:HERA-F2-q2bins}
\end{figure}

The region around $x=0.25$, in which scaling was observed at SLAC, 
can be seen in figure~\ref{fig:HERA-F2-xbins}.
Despite the fact that the data~\cite{H1F2-9697a,H1F2-9697b,ZEUSF2-9697} now extend over four 
orders of magnitude 
in $Q^2$, approximate scaling is still observed. However, if one examines
the data over the much wider range of $x$ available to the HERA experiments, 
it is clear that scaling is badly violated at low $x$. This rapid
rise of the structure function at low $x$ (see figure~\ref{fig:HERA-F2-q2bins})
can be attributed to
gluon emission and the subsequent production of virtual quark-antiquark
pairs, which can in their turn radiate gluons, producing a ``sea'' of 
partons at lower and lower $x$. Thus, the precise measurement of the proton
structure at low $x$ at HERA is very sensitive both to the details 
of the evolution in QCD of this shower of partons and to the
value of the strong coupling constant, $\alpha_s$. 

\begin{figure}[h]
\begin{center}
\epsfig{file=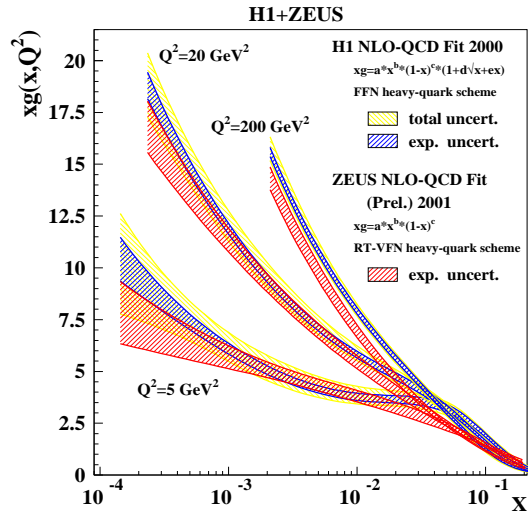,%
      width=7cm,%
      height=7cm%
        }
\end{center}
\caption{The gluon density in the proton as measured by ZEUS (red shaded
band) and H1~\protect{\cite{H1F2-9697b}} (yellow and blue shaded bands) as a function
of $x$ in three bins of $Q^2$. The functional form used by the two
collaborations in the gluon fit is shown in the legend.} 
\label{fig:gluon}
\end{figure}

The sensitivity of the evolution of $F_2$ to the value of $\alpha_s$ has
been exploited by both ZEUS and H1. Each experiment has made a global QCD fit to
its own data plus some or all of the fixed-target DIS data. There is
reasonably good agreement between the experiments,
although work still continues on understanding the differing treatments
of the errors in the two experiments. The results are
shown in figure~\ref{fig:gluon}. 

\begin{figure}[h]
\begin{center}
\epsfig{file=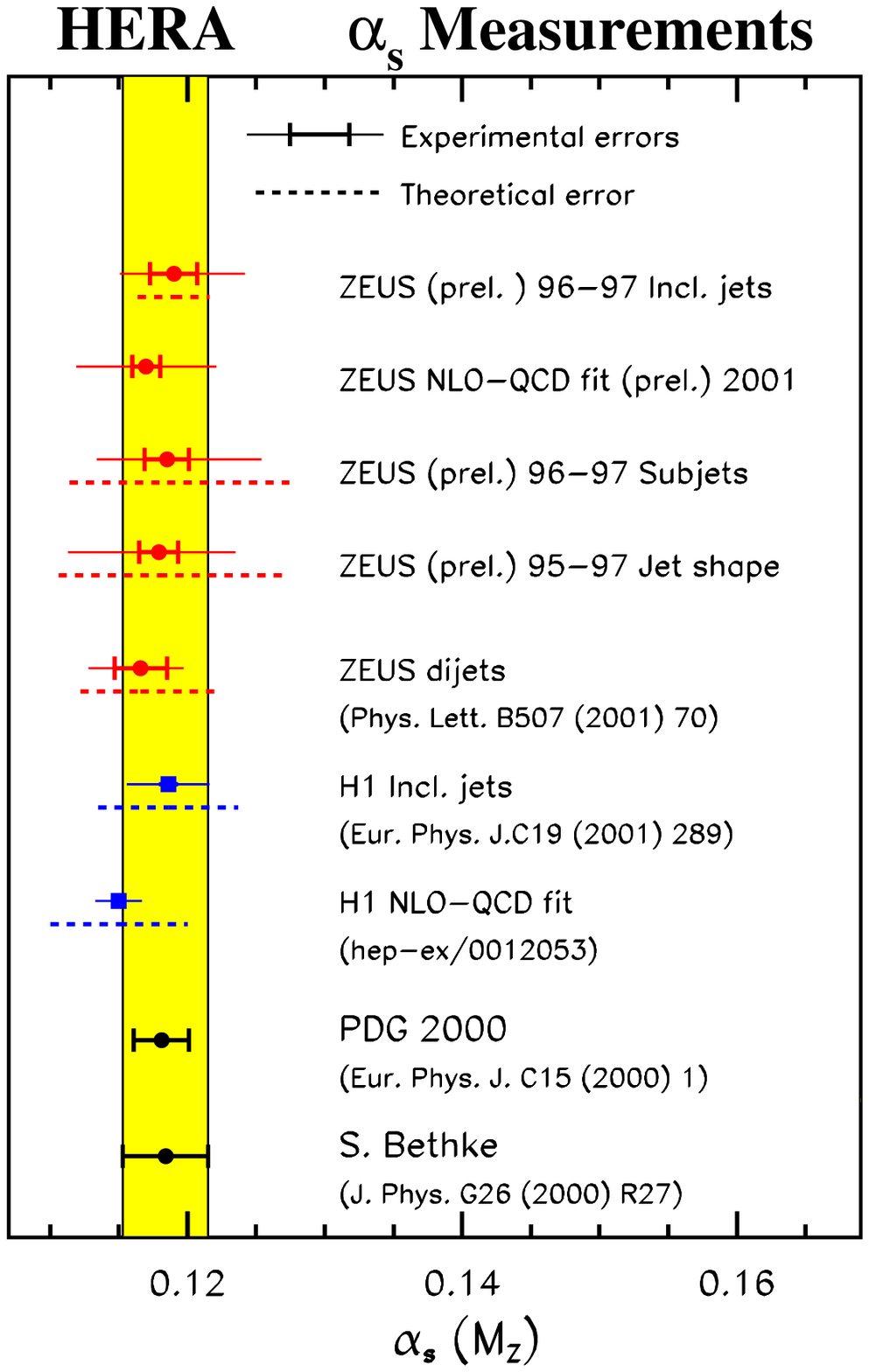,%
      width=9cm,%
      height=9cm%
        }
\end{center}
\caption{Values of the strong coupling constant as determined at HERA. Each
different measurement is displaced vertically for ease of visibility;
each value arises from a different method as briefly indicated in the
legend. The reference for published results is shown below the method
label. The world
average as calculated by the particle data group and by Bethke are
shown at the bottom of the figure.}
\label{fig:alphas}
\end{figure}

Another output of the fit is a
value of $\alpha_s$; the results from the two experiments are shown in
figure~\ref{fig:alphas}, labelled as ``NLO-QCD fit''. Also shown are a variety
of other, high-precision, measurements of $\alpha_s$ that can be
made at HERA using a variety of techniques. These include classic methods
such as the rate of dijet + proton-remnant production compared to that
of single jet plus remnant, the subjet-multiplicity evolution
inside jets and the shape of jets. Many of these give excellent
precision, comparable to the world average. The dominant uncertainty
is usually theoretical and arises from the lack of next-to-next-to-leading-order 
predictions. 

\begin{figure}[h]
\begin{center}
\epsfig{file=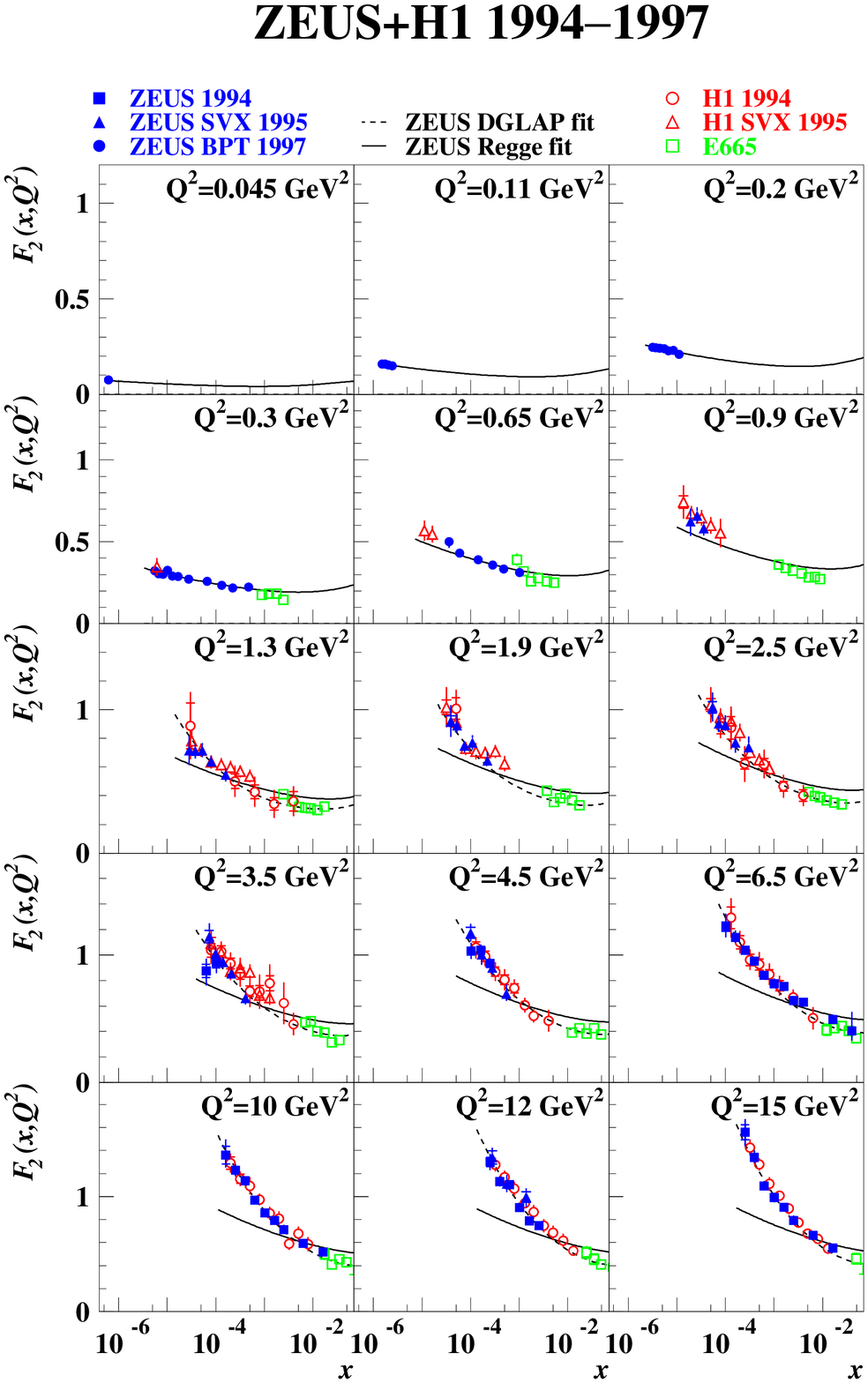,%
      width=9cm,%
      height=11.5cm%
        }
\end{center}
\caption{ZEUS BPT data on $F_2$ in bins of $Q^2$ as
a function of $x$. Also shown are earlier
ZEUS data as well as data from H1 and E665.
The solid line shows the results of the ``ZEUS Regge fit" to
the form of equation~\protect\ref{eq:GVDM+REGGE}, while 
the dotted line shows the result of the ZEUS NLO QCD fit.} 
\label{fig:ZEUSBPT:q2bins}
\end{figure}

The structure-function data are not only sensitive to QCD effects. The publication of the ZEUS ``BPT'' data~\cite{ZEUS-BPT} gives access
to very low $Q^2$ and $x$ regimes. As shown by figure~\ref{fig:ZEUSBPT:q2bins}, 
although QCD gives a good fit to the data down to $Q^2 \sim 1$ GeV$^2$,
below that it is necessary to use a Regge-based fit of 
the form 

\begin{equation}
F_2(x,Q^2) = \left(\frac{Q^2}{4\pi^2\alpha} \right)
\left(  \frac{M_0^2}{M_0^2+Q^2}\right) \left( A_\reg
\left( \frac{Q^2}{x} \right)^{\alpha_\reg-1}+A_\pom
\left(\frac{Q^2}{x}\right)^{\alpha_\pom-1}\right),
\label{eq:GVDM+REGGE}
\end{equation}
where $A_\reg, A_\pom$ and $M_0$ are constants and $\alpha_\reg$ and
$\alpha_\pom$ are the Reggeon and Pomeron intercepts, respectively.
Regge theory is expected to apply at asymptotic energies. The
appropriate energy here is $W$, the
centre-of-mass energy of the virtual photon-proton system, which can
be expressed in terms of the other invariants as
\begin{equation}
W^2 = Q^2\frac{1-x}{x} . 
\label{eq:w2q2x}
\end{equation} 
Since, at low $x$, $W^2 \sim 1/x$, it would be expected that Regge fits would be 
applicable at very low $x$ and $Q^2$. 

\begin{figure}[h]
\begin{center}
\epsfig{file=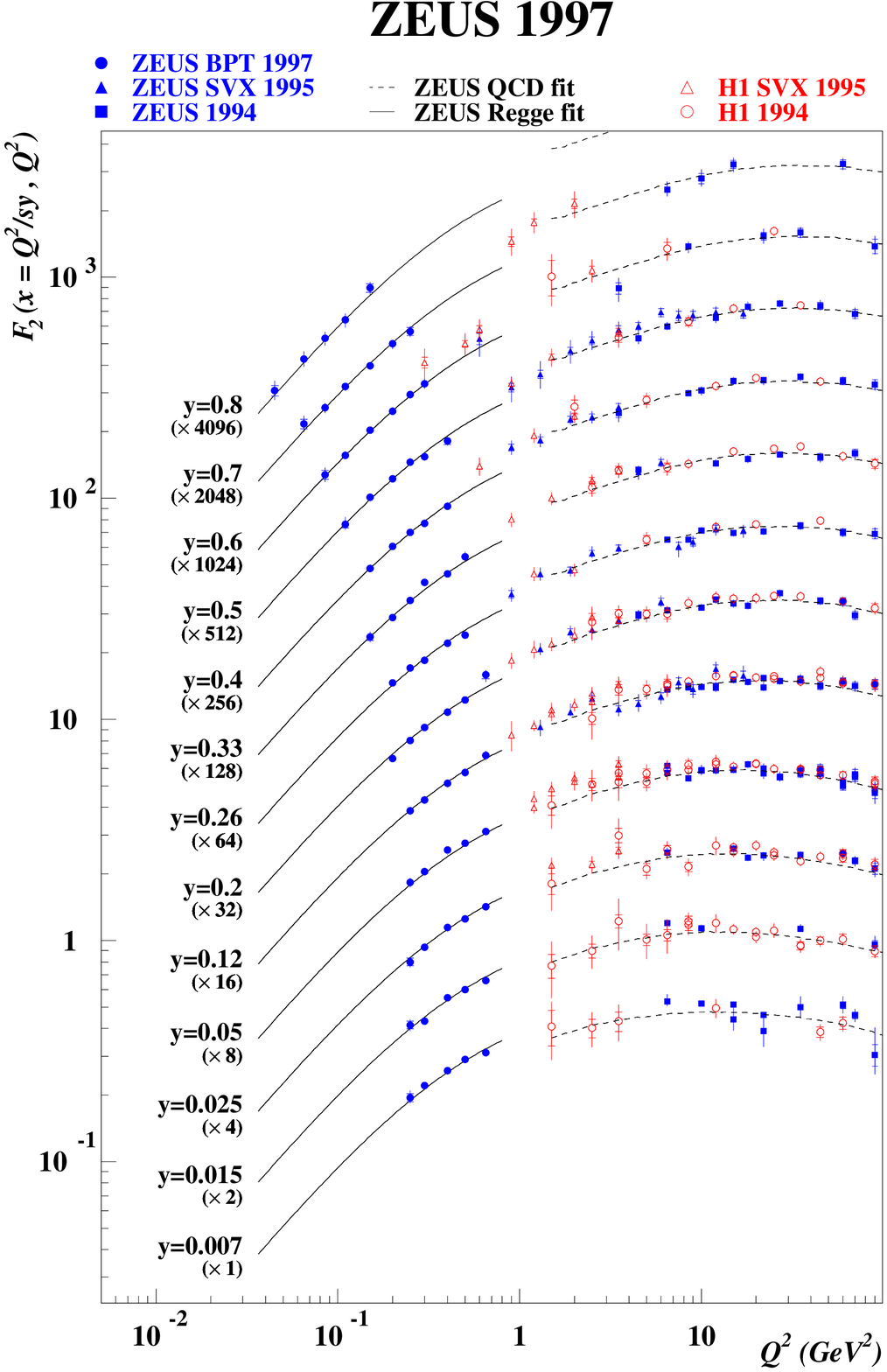,%
      width=10cm,%
      height=10cm%
        }
\end{center}
\caption{ZEUS BPT data on $F_2$ in bins of $y$ as
a function of $Q^2$. Also shown are earlier
ZEUS data as well as data from H1.
The solid line shows the results of the ``ZEUS Regge fit" to
the form of equation~\protect\ref{eq:GVDM+REGGE}, while
the dotted line shows the result of the ZEUS NLO QCD fit.}
\label{fig:ZEUSBPT:ybins}
\end{figure}
Figure~\ref{fig:ZEUSBPT:ybins} shows the ZEUS BPT data,
together with both ZEUS and H1 $F_2$ data, in bins
of constant $y$ as a function of $\ln Q^2$. 
For $Q^2~\gsim~1$
GeV$^2$, the data are roughly independent of $Q^2$, whereas at lower $Q^2$
they fall rapidly, approaching the $Q^{-2}$ dependence that would be expected in
the limit $Q^2 \rightarrow 0$ from conservation of the electromagnetic current. 

\begin{figure}[t]
\begin{center}
\epsfig{file=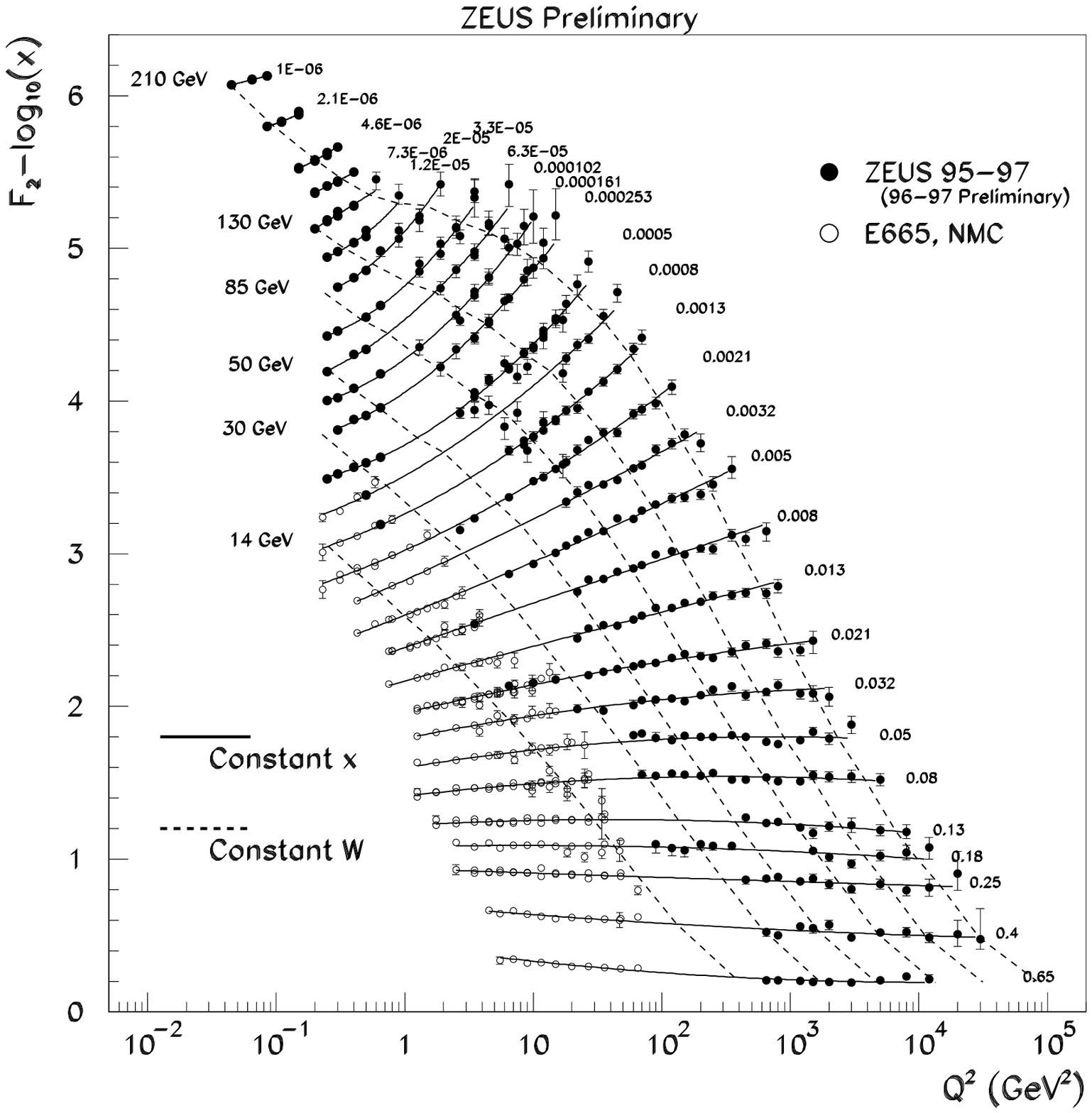,%
      width=11cm,%
      height=11cm%
        }
\end{center}
\caption{Compilation of ZEUS $F_2$ data in $x$ bins as a function of $Q^2$. 
Each $x$ bin is shifted by an additive constant for ease of visibility.
Data from NMC and E665 are also shown.
The dotted lines show lines of constant $W$, while the solid lines are
fits to the form of equation~\protect\ref{eq:F2param}.} 
\label{fig:ZEUS:6OF2}
\end{figure}

The combination of the BPT data and the latest
$F_2$ data means that ZEUS now has precise data over a 
remarkable six orders of magnitude in $x$ and $Q^2$. 
These data are shown in $x$ bins as a function of $\ln Q^2$ 
in figure~\ref{fig:ZEUS:6OF2}, together with fixed target data
from NMC and E665, which extends the range in the direction of medium
$x$ and $Q^2$. 

The availability of this very wide range of precise data makes possible qualitatively 
new investigations of models that describe $F_2$. Since the logarithmic derivative of 
$F_2$ is directly
proportional to the gluon density in leading-order QCD, which in turn is 
the dominant parton density at small $x$, its behaviour as a function of both
$x$ and $Q^2$ is important. The solid curves
on the figure correspond to fits to a polynomial in $\ln Q^2$ of the form
\begin{equation}
F_2 = A(x) + B(x) \left(\log_{10} Q^2\right) + C(x) 
\left(\log_{10}Q^2\right)^2,
\label{eq:F2param}
\end{equation}
which gives a good fit to the data through the entire kinematic range. The dotted lines on figure~\ref{fig:ZEUS:6OF2} are lines of constant $W$. 
The curious `bulging' shape of these contours in the
small-$x$ region immediately implies that something interesting is
going on there. Indeed, simple inspection of figure~\ref{fig:ZEUS:6OF2} shows
that the slope of $F_2$ at constant $W$ begins flat in the scaling region,
increases markedly as the gluon grows and drives the evolution of $F_2$
and then flattens off again at the lowest $x$. 

\begin{figure}[h]
\begin{center}
\epsfig{file=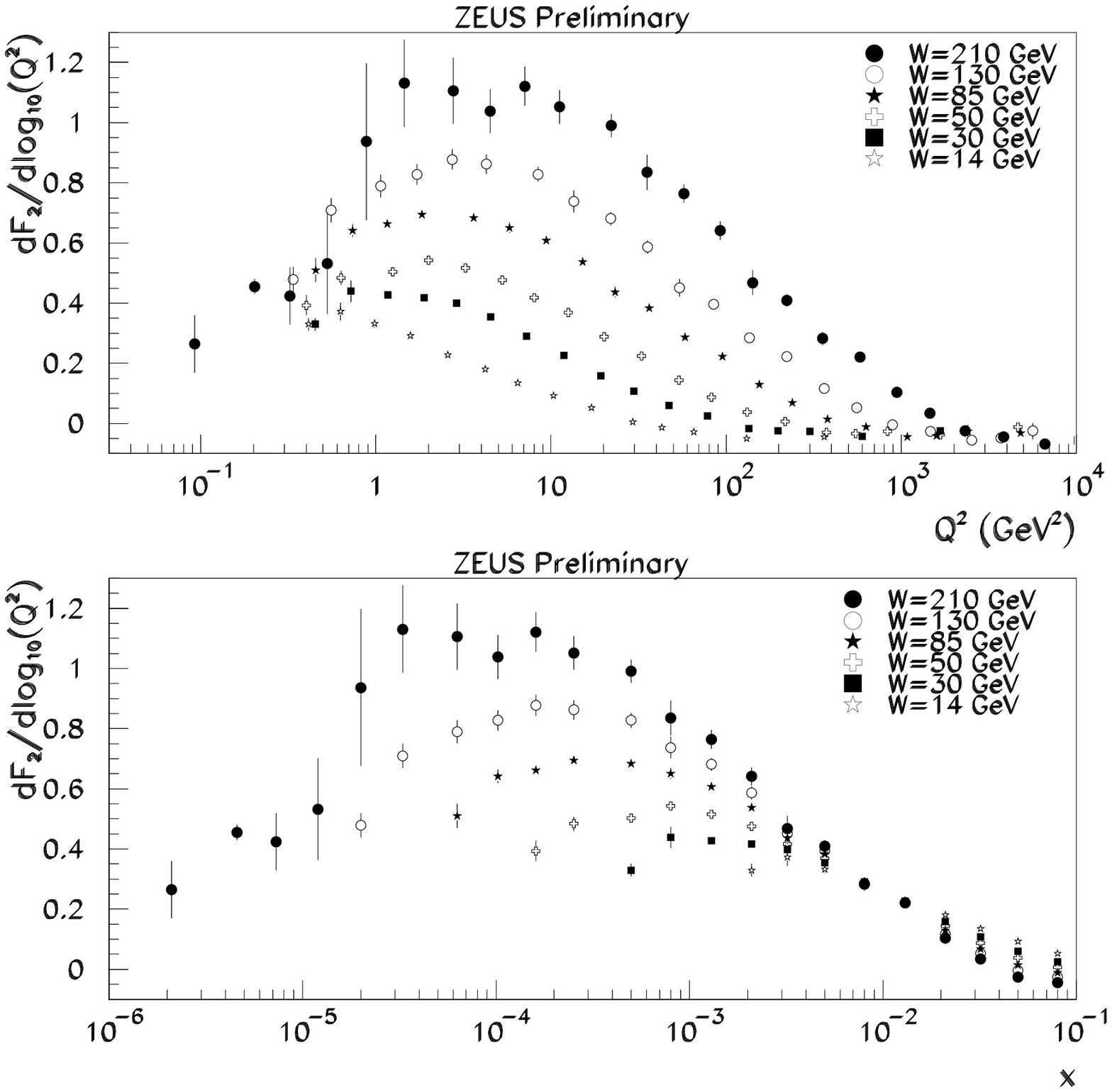,%
      width=9cm,%
      height=9cm%
        }
\end{center}
\caption{The logarithmic derivative of the ZEUS $F_2$ data in six bins
of $W$, plotted as a function of $Q^2$ and $x$.}
\label{fig:ZEUS:logder}
\end{figure}

Figure~\ref{fig:ZEUS:logder}
shows the logarithmic derivative evaluated at $(x, Q^2)$ points
along the contours of fixed $W$
shown on figure~\ref{fig:ZEUS:6OF2} according to the derivative of
equation~\ref{eq:F2param}, {\it viz.}:
\begin{equation}
\frac{\partial F_2}{\partial \log_{10} Q^2} = B(x) + 2C(x) \log_{10} Q^2,  
\label{eq:F2deriv}
\end{equation}
where the data are plotted separately as functions of $\ln Q^2$ and $\ln x$.
The turn-over in the derivatives in all $W$ bins is marked. Within the framework of 
pQCD, the interpretation of such an effect is that the growth of the gluon 
density at low $x$ is tamed as $Q^2$ and $x$ fall. Such an effect
is by no means necessarily an indication of deviations from the 
standard DGLAP~\cite{sovjnp:20:95,sovjnp:15:438,np:b126:298,jetp:46:641}
evolution. Nevertheless, such a fall in the gluon 
density as $x$
falls is a natural consequence of parton saturation or shadowing. 
These effects can be naturally discussed
in ``dipole models''~\cite{BFrsreview}, which often explicitly
take into account parton-saturation effects. 
In such models, the ``standard'' picture of deep 
inelastic scattering in the infinite-momentum frame of the proton is
replaced by an equivalent picture produced by a Lorentz boost into
the proton rest frame. In this frame, the virtual photon undergoes
time dilation and develops structure far downstream of the interaction
with the proton. The dominant configurations of this structure are
$q\overline{q}$ and $q\overline{q}g$ Fock states, which interact with
the proton as a colour dipole. The higher the $Q^2$ of the interaction,
the smaller the transverse size of the dipole. For small $x$, the deep inelastic process 
can be considered semi-classically as the coherent interaction of the dipole with the
stationary colour field of the proton a long time after
the formation of the dipole. 

\begin{figure}[h]
\begin{center}
\epsfig{file=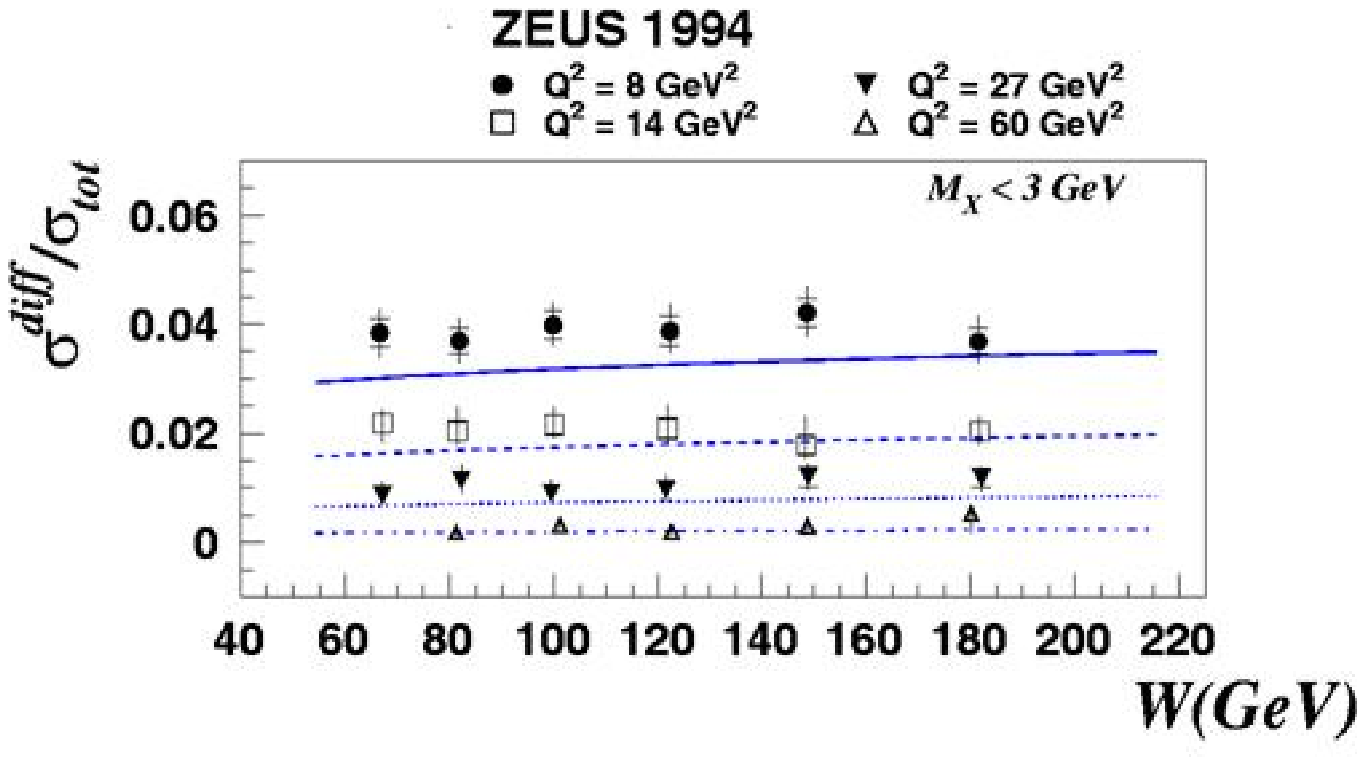,%
      width=11cm,%
      height=8cm%
        }
\end{center}
\caption{The ratio of the diffractive to total cross section in four $Q^2$ bins as a 
function of $W$. The curves show the predictions from the
Golec-Biernat \& Wusthoff model.}
\label{fig:diff-tot-ratio}
\end{figure}

One of the most attractive features of such models is 
the rather natural way in which they can lead to a
unified description of diffraction and 
deep inelastic scattering. Diffractive DIS
is a subset of fully inclusive DIS characterised by a hard interaction between the proton and the exchanged virtual photon that nevertheless leaves the proton intact. The fully inclusive structure functions
sum over all possible exchanges between the dipole and the proton, dominantly
one- and two-gluon exchange in a colour octet, whereas diffraction is produced by the exchange of two gluons in a colour-singlet state. This deep connection between these two processes leads to non-trivial
predictions~\cite{pr:d59:014017,pr:d60:114023} which do indeed seem to be at 
least qualitatively in agreement
with the data. 
This is illustrated in figure~\ref{fig:diff-tot-ratio}.
This figure is surprising for several reasons. It demonstrates that the diffractive cross section
has the same $W$ dependence as the total cross section. To the extent to
which the diffractive cross section can be related to the elastic cross section,
one would have expected from the Optical Theorem that the
ratio would have a power-law dependence on $W$, as indeed would also be
expected from Regge theory via the exchange of a Pomeron. A strong $W$ 
($\sim 1/x$) dependence is also expected in QCD models, since the total cross section is dominated by single-gluon
exchange, whereas diffraction is dominated by
two-gluon exchange. The other surprise is the fact that the GBW model gives
a rather good qualitative representation of the data. The behaviour of this
ratio as $Q^2 \rightarrow 0$ is also likely to be of great interest. In this
talk, however, there is only time to scratch the surface of these interesting
low-$x$ and diffractive phenomena, which contain a great deal of
information touching on the very challenging problem of confinement in QCD. 

\subsection{High-$Q^2$ phenomena}
\label{sec:highQ2}

\begin{figure}[h]
\begin{center}
\epsfig{file=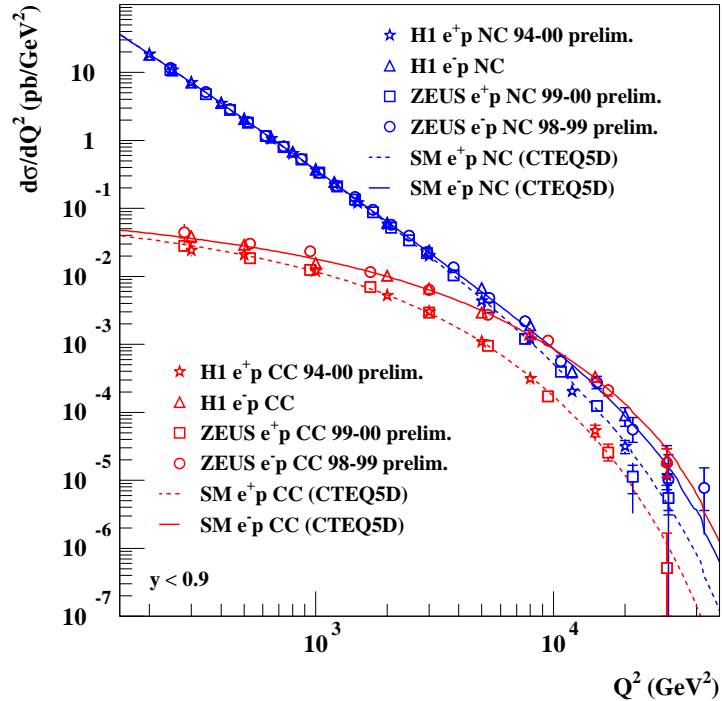,%
      width=11cm,%
      height=11cm%
        }
\end{center}
\caption{Charged and neutral current differential cross sections for
$e^\pm$ scattering as a 
function of $Q^2$ from H1 and ZEUS.}
\label{fig-H1ZEUS-CCNC}
\end{figure}

HERA provides an unique opportunity to study the electroweak 
interaction at $Q^2$ sufficiently high that the charged and neutral 
currents are of similar strength. 
Figure~\ref{fig-H1ZEUS-CCNC} shows the differential cross-sections for the 
charged and neutral currents as a function of $Q^2$ from H1 and ZEUS. 
It can be seen that, for $e^-p$ interactions, 
these two processes become of equal strength at $Q^2$ 
$\sim M_Z^2 \sim 10^4$ GeV$^2$. For $e^+p$ interactions, 
the charged current cross-section approaches the neutral current 
cross-section, but remains below it. The features of
this plot can be explained  
by inspection of equation~\ref{eq:Fl:sigma}, together with
\ref{eq-CCem} and \ref{eq-CCep} below:
\begin{eqnarray} 
\left. \frac{d^2 \sigma}{dx dQ^2} \right|_{\emin}^{CC} & = &   
\frac{G_F^2}{2\pi} \left( \frac{\MWtwo}{\MWtwo+Q^2}\right)^2 \cdot
 \nonumber \\ 
 & & 2x \{u(x) + c(x) + (1-y)^2(\overline{d}(x) + \overline{s}(x))\} 
\label{eq-CCem}  
\end{eqnarray} 

\begin{eqnarray} 
\left. \frac{d^2 \sigma}{dx dQ^2} \right|_{\eplus}^{CC} & = &   
\frac{G_F^2}{2\pi} \left(\frac{\MWtwo}{\MWtwo+Q^2}\right)^2 \cdot
\nonumber \\ 
 & & 2x\{\overline{u}(x) +\overline{c}(x) + (1-y)^2(d(x) + s(x))\}
\label{eq-CCep} 
\end{eqnarray}
For the charged current case, the smaller 
size of the $e^+p$ cross-section compared to $e^-p$ is related to the
fact that, at high $Q^2$, equation~\ref{eq:yq2sx}
implies that both $ x, y \rightarrow  
1$.  There are two main contributory factors to the cross-section
difference that flow from this. First, there are twice as many $u$ valence quarks inside the proton that can couple to $W^-$ as $d$ quarks that
can couple to $W^+$. Secondly, 
the $(1-y)^2$ terms in equations~\ref{eq-CCem} and \ref{eq-CCep}, which 
arise from the $V - A$ helicity structure of the charged weak current, 
imply that the valence-quark contribution, which is dominant at high $Q^2$,
is suppressed for the positron case but not for electrons.

\begin{figure}[h]
\begin{center}
\epsfig{file=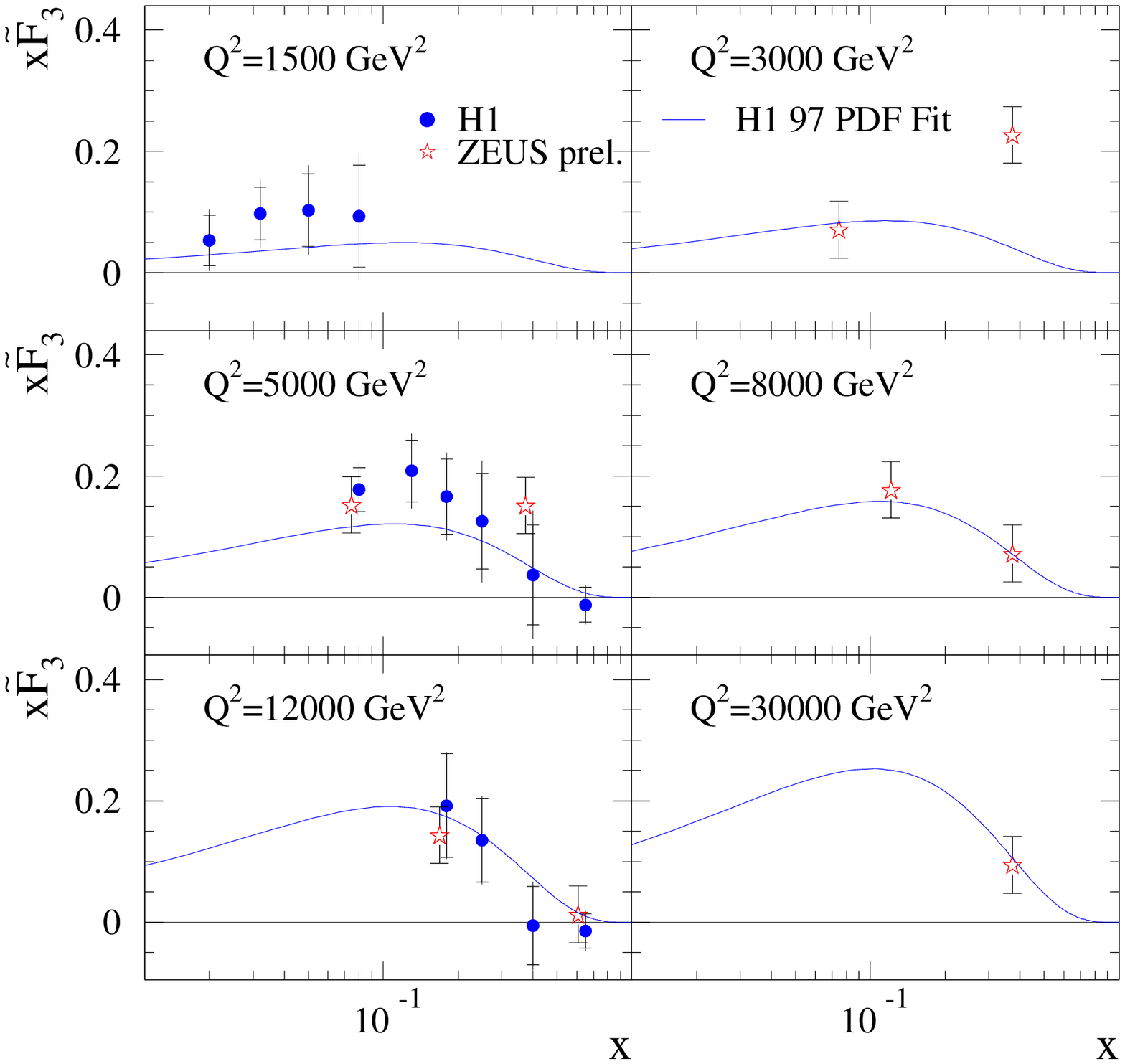,%
      width=11cm,%
      height=11cm%
        }
\end{center}
\caption{The $xF_3$ structure function as determined by
H1 and ZEUS as a function of $x$ in
six bins of $Q^2$.}
\label{fig-H1ZEUS-xF3}
\end{figure}

The difference between the electron and positron neutral current
cross sections shown in equation~\ref{eq:Fl:sigma} allows the determination
of the parity-violating structure function $xF_3$ by taking the difference
of the cross sections. The results~\cite{H1-xF3} are shown in 
figure~\ref{fig-H1ZEUS-xF3}. This is clearly a very difficult measurement since it requires the subtraction of
two quantities that are almost equal. The measurement is dominated by
statistical errors and particularly by the fact that the electron data sample that has so far been obtained at HERA is much smaller than that for positrons.

\begin{figure}[h]
\begin{center}
\epsfig{file=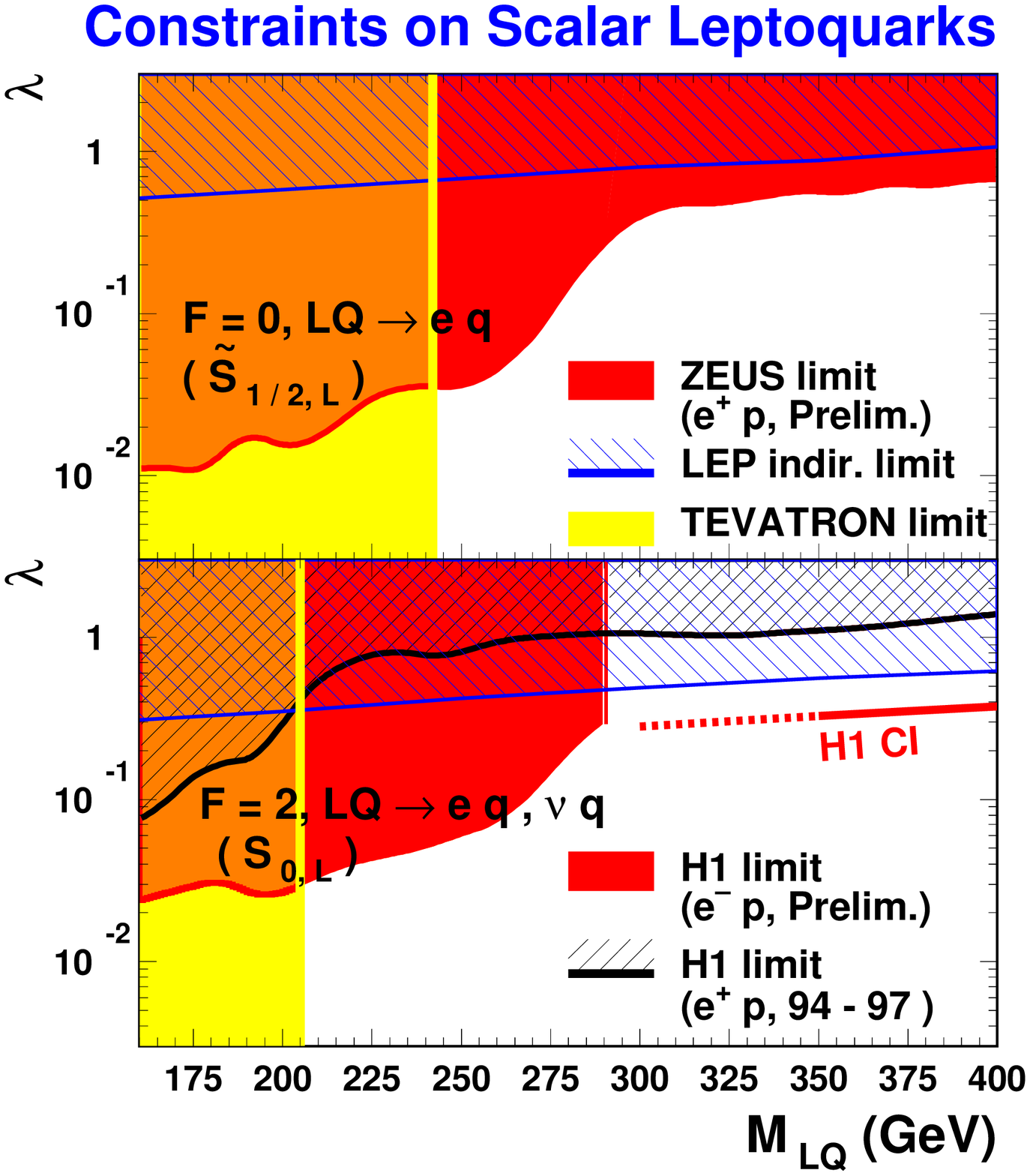,%
      width=10cm,%
      height=10cm%
        }
\end{center}
\caption{Limits on coupling strength $\lambda$ versus mass $M_{LQ}$  
for leptoquarks. The top plot shows limits
for fermion number = 0 leptoquarks decaying into the $eq$ final state from ZEUS. The lower plot shows limits from H1 for fermion 
number = 2 leptoquarks decaying into both $eq$ and $\nu q$ final states.
Also shown
are limits obtained from the Tevatron (yellow shaded area) and LEP
(blue striped area). These leptoquark species have identical quantum numbers
to squarks that violate R-parity.}
\label{fig-ZEUSH1-LQ}
\end{figure}

The high-$Q^2$ regime is also interesting since possible
new states from electron-quark fusion (e.g.\
leptoquarks) have masses given by $M^2 \sim sx$ and since the sensitivity
to the effects of new currents is maximised. An example of the
sensitivity that can be obtained at HERA is shown in figure~\ref{fig-ZEUSH1-LQ},
which shows the mass against coupling limits for two varieties of scalar leptoquark. Both H1 and ZEUS have comparable 
limits for a whole range of
such states with differing quantum numbers. It can be seen from
figure~\ref{fig-ZEUSH1-LQ}, and it is generally the case, that
for some states, in particular in $R$-parity violating supersymmetry
models or leptoquarks, HERA has higher sensitivity than either LEP or the
Tevatron.  

\begin{figure}[h]
\begin{center}
\epsfig{file=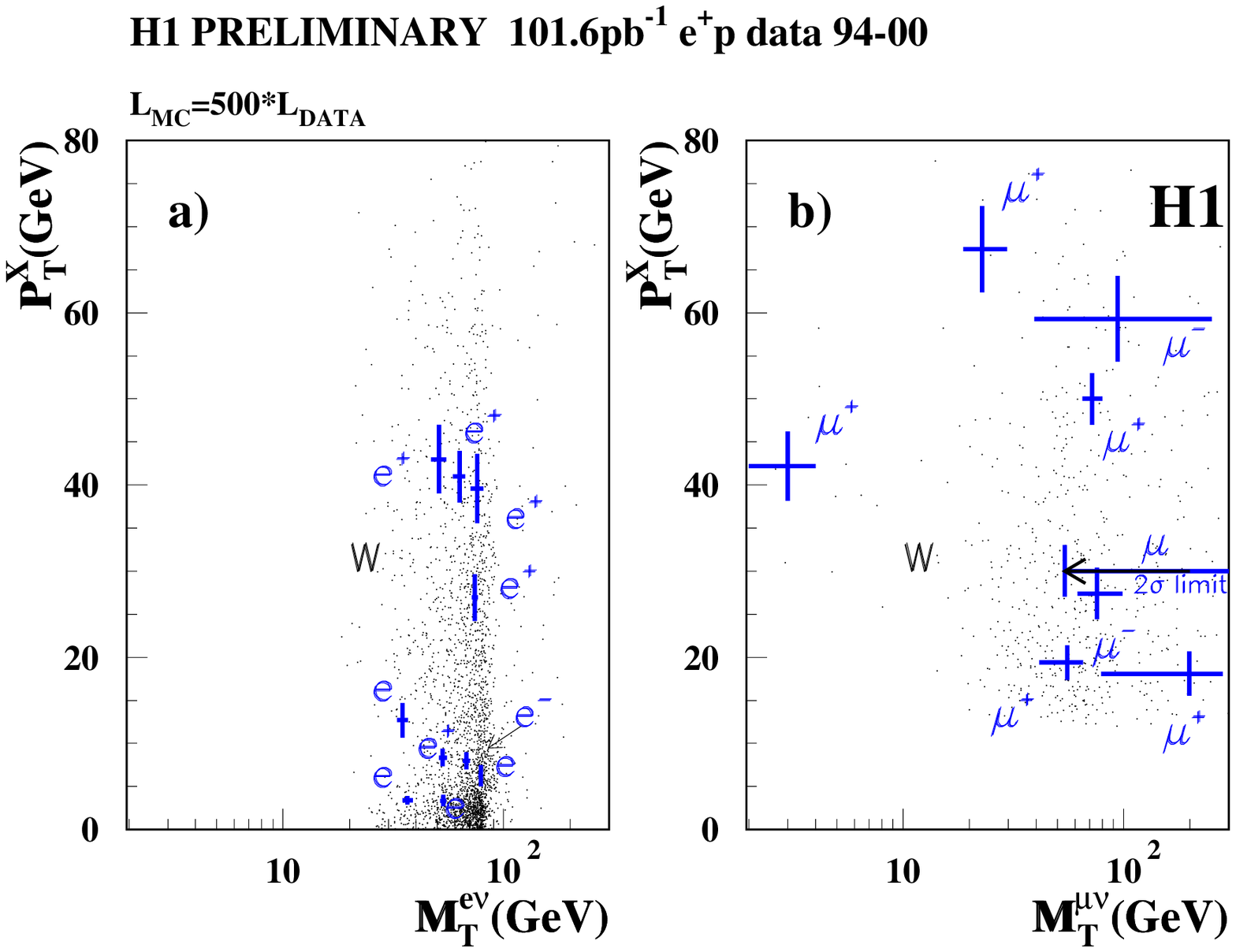,%
      width=11cm,%
      height=9cm%
        }
\end{center}
\caption{Distribution of transverse mass versus the $p_T$ of the
hadronic system for H1 events containing a) isolated electrons and b) isolated
muons. The dots show the distribution of Standard Model
$W$ Monte Carlo events corresponding to a 
luminosity 500 times that of the data.}
\label{fig-H1leptons}
\end{figure}

As well as stringent limits on new phenomena, the HERA data also show
intriguing features which may be signatures for new physics. 
The H1 collaboration has observed a class of
events that have isolated charged leptons with
large missing transverse momentum. Figure~\ref{fig-H1leptons} shows the distribution of the transverse
momentum of the hadronic system, $p_T^X$, against its transverse mass 
separately for electrons (or positrons) and muons in such events. Also shown are the expectations from the Standard Model background, which is dominated by
single $W$ production. 

It can be seen that the distribution of the events
is rather different to the Standard Model expectation. Furthermore,
H1 sees an excess of such events. For the transverse mass of the
hadronic system greater than 25 GeV, H1 sees four electron and 
six muon events, compared to Standard Model expectations of
1.3 and 1.5 events, respectively. Unfortunately, this exciting
observation is not confirmed by ZEUS, which, for the same cut in
$p_T^X$, sees one event in each category compared to the Standard
Model expectation
of 1.1 and 1.3, respectively. Intensive discussions between the two
experiments have not revealed any reason why H1 might artificially
produce such an excess nor why ZEUS should not observe it. It would therefore seem that there must be an unlikely fluctuation: either the H1
observation is an upward fluctuation from the Standard Model, or
ZEUS has suffered a downward fluctuation from a signal for new physics. 
More data from HERA II will be required to resolve this puzzle.

\begin{figure}[h]
\begin{center}
\epsfig{file=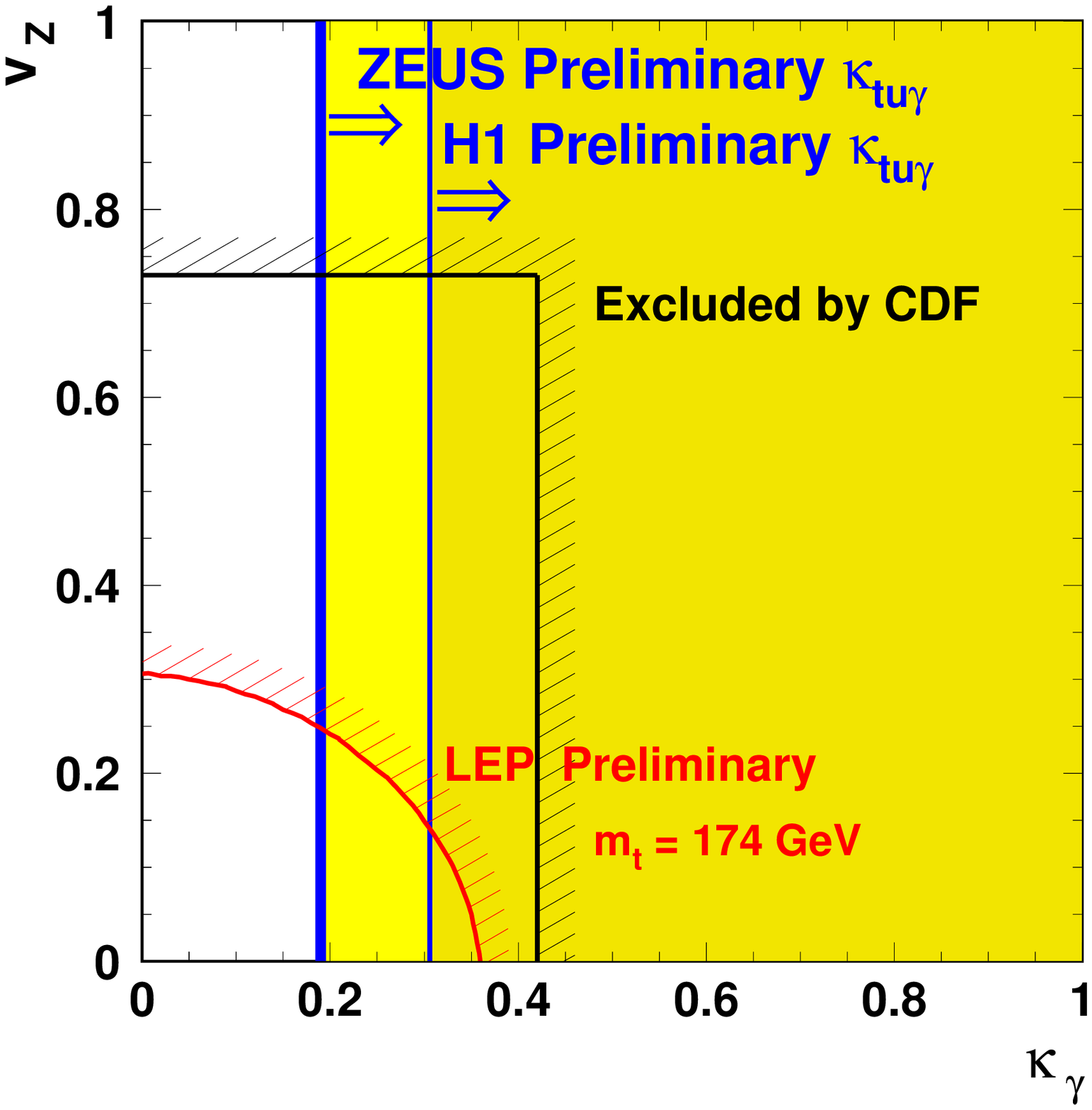,%
      width=9cm,%
      height=9cm%
        }
\end{center}
\caption{Limits on flavour-changing neutral current coupling strength for
single top production. Limits from H1 and ZEUS are plotted 
on the two-dimensional space of the photon and $Z$ coupling strengths.
Also shown are similar limits from the combined LEP experiments 
(to the right of the shaded curve) and 
from CDF (to the right of the black shaded lines).}
\label{fig-H1ZEUS-FCNC}
\end{figure}

One possible source of an excess of events with isolated leptons with
missing transverse momentum would be from a flavour-changing neutral
current process producing single top quarks. Both H1 and ZEUS have
used the samples described above to put limits on the FCNC couplings
of the $\gamma$ to light quark-top quark vertices. The
results are shown in figure~\ref{fig-H1ZEUS-FCNC}. Also shown are
the limits from LEP and CDF, which are complementary to those
from HERA, in the sense that, since the $Z$-exchange cross section at
HERA is so much smaller than that for $\gamma$ exchange, the HERA data limit only the photon coupling.

\section{HERA II physics}
\label{sec:HERAII}

Many of the physics results discussed above, particularly those at
high $Q^2$, are statistics limited. Moreover, there is a natural build-up of
transverse polarisation of the lepton beam in HERA that occurs through the Sokholov-Ternov effect~\cite{S-Teffect}. As very successfully
demonstrated at HERMES using gas targets, this transverse polarisation can be
rotated into the longitudinal direction and utilised to do physics. The
installation of spin rotators in H1 and ZEUS would allow polarisation
studies to be carried out at very much higher $Q^2$. This is particularly
interesting to study the chiral properties of the electroweak interaction.
For these and several other reasons, it was decided to embark on a major
upgrade of both the HERA accelerator and the H1 and ZEUS detectors. The
aim of the HERA II programme is to produce a factor of approximately five increase in luminosity and accumulate 1 fb$^{-1}$ of data with 
both electron and positron collisions in both longitudinal polarisation states.

The changes to the HERA accelerator include the replacement of 480 meters of the vacuum system and the design and installation of almost 80 magnets is the region around the H1 and ZEUS interaction points. In particular, superconducting
quadrupole focussing elements were inserted inside both detectors to reduce
the beam emittance and spin rotators were installed on either side
of the H1 and ZEUS interaction regions. 

Both the ZEUS and H1 detectors have undergone a massive programme of 
consolidation and repair work, as well as major detector upgrades. In the time
available I can only discuss briefly the changes made to ZEUS; the general
thrust of the upgrade is similar in the two detectors, but the details are different.

\subsection{Upgrades to ZEUS for HERA II}
\label{sec:ZEUSupgrade}
The ZEUS upgrades have concentrated in three main areas: the vertex region;
the forward (= proton beam) direction; and the luminosity monitoring.

\subsubsection{The vertex region}
\label{sec:ZEUSvertex}

\begin{figure}[h]
\begin{center}
\epsfig{file=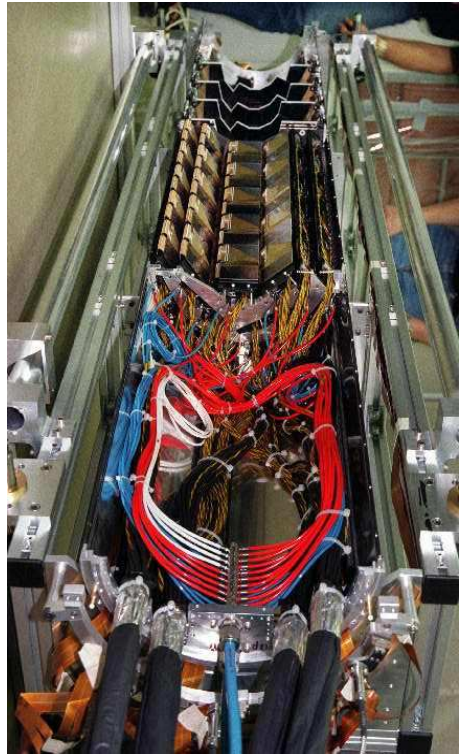,%
      width=6cm,%
      height=10cm%
        }
\end{center}
\caption{A photograph of one half of the MVD, showing the barrel ladders,
one half of each of the four forward wheels and the cables and services.}
\label{fig:MVD-barrel-photo}
\end{figure}

The tagging of the large flux of heavy quarks (charm and beauty) produced at HERA II can be greatly enhanced by the installation of a high-precision
charged-particle detector as close as possible to a thin beampipe. 
The ZEUS MVD~\cite{nim:a435:34,nim:a473:26} 
consists of 20 $\mu$m pitch $n$-type silicon-strip detectors 
with $p^+$-type implants. The readout pitch is 120 $\mu$m, leading
to more than 200,000 readout channels, which are digitised by a
custom-built clock, control and ADC system. 
The detectors are organised in two main groups: a ``barrel'', which surrounds the elliptical 2 mm-thick ($\sim 1.1$\% of a radiation length) 
aluminium-beryllium beam-pipe; and four ``wheels'', consisting
of wedge-shaped detectors mounted 
perpendicular to the beam-line in the forward direction from the
interaction point. 
Figure~\ref{fig:MVD-barrel-photo} shows one half
of the MVD before installation at DESY. In the barrel region, the ladders, each of which consists of five silicon detectors, and halves of the
four forward ``wheels'', can be seen, as can the dense array of readout and
services cables and the cooling system. The complete MVD was installed in ZEUS
in April 2001 and has been fully integrated with the ZEUS DAQ system;
both cosmic-ray and beam-related data have been taken. 

\begin{figure}[h]
\begin{center}
\epsfig{file=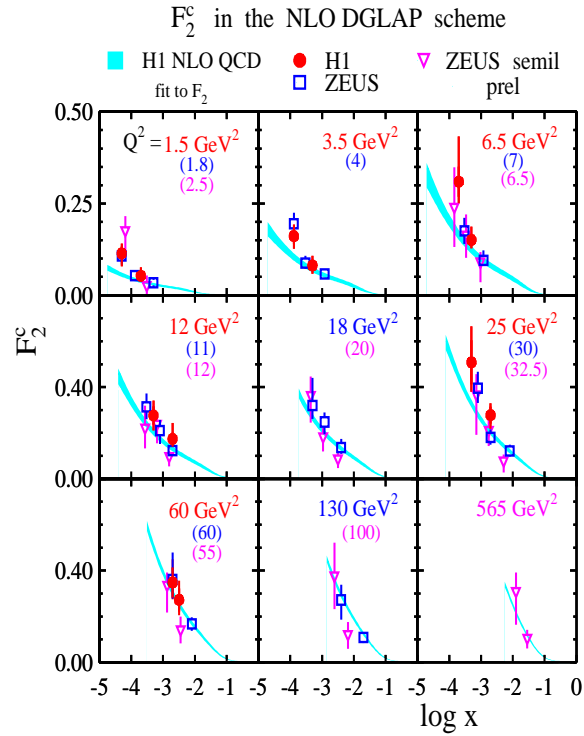,%
      width=8cm,%
      height=10cm%
        }
\end{center}
\caption{Values of the charm structure function, $F_2^c$ from the H1
and ZEUS experiments in bins of $Q^2$ as a function of $\ln x$. The 
shaded curves show the predictions from the NLO QCD fit to the inclusive
$F_2$ data by H1.}
\label{fig-H1ZEUS-F2C}
\end{figure}

\begin{figure}[h]
\begin{center}
\epsfig{file=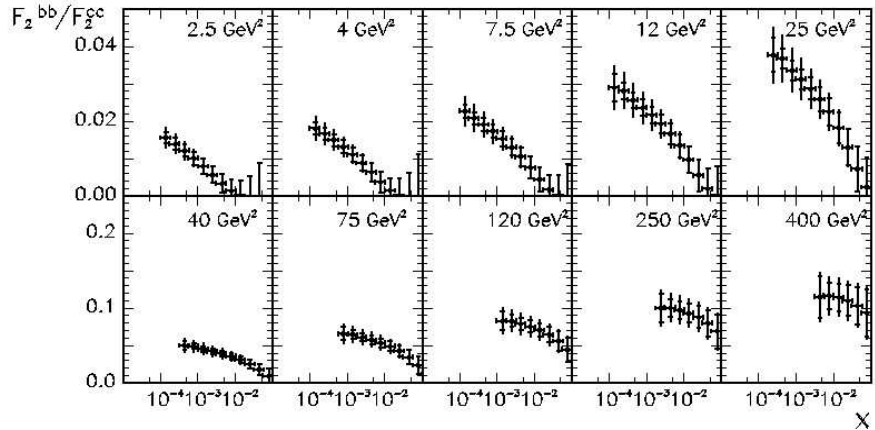,%
      width=12cm,%
      height=6cm%
        }
\end{center}
\caption{The MC prediction for the ratio of the contribution to $F_2$
of $b$-quark to $c$-quark production
in $Q^2$ bins as a function of $x$ after 500 pb$^{-1}$ of data at
HERA II.}
\label{fig:F2b-c-ratio}
\end{figure}

The physics programme addressed by the MVD is that of the flavour decomposition of the proton and photon and the search for physics beyond the Standard Model.
The measurements of the semi-inclusive charm structure function, $F_2^c$, made
by both experiments~\cite{ZEUS-F2c,H1-F2c} are shown in figure~\ref{fig-H1ZEUS-F2C}. 
The large increase in luminosity of HERA II, together with the ability to
tag heavy-quark decays in the MVD, should greatly improve 
the measurement of $F_2^{c}$. After about 500~pb$^{-1}$, 
an uncertainty of around the 2\% currently measured on $F_2$ should
be obtained. In addition, $b$-quark production can be measured precisely; a
Monte Carlo simulation~\cite{proc:hera:1995:89} of a measurement of
$F_2^{b}/F_2^c$ after 500 
pb$^{-1}$ is shown in figure~\ref{fig:F2b-c-ratio}. 

It should also be possible, from a combination of neutral and charged
current measurements, to separate out the $u,d,s,c,b$ and $g$ contribution to $F_2$. 

\subsubsection{Charged-particle tracking in the forward direction}
\label{sec-stt}

The higher luminosity expected at HERA II will increase the number of very
high-$Q^2$ events in which the electron or positron is scattered into the forward direction. It will also give access to rare processes, including
possible physics beyond the Standard Model, which tend to have 
forward jets and/or leptons. The pattern-recognition capabilities of the ZEUS Forward
Tracker have therefore been improved by the replacement of two layers of transition-radiation detector by layers of straw tubes. The straws are approximately 7.5 mm in diameter and
range in length from around 20 cm to just over 1 m. They are constructed from
two layers of 50 $\mu$m kapton foil coated with a 0.2 $\mu$m layer of aluminium, surrounding a 50 $\mu$m wire
at the centre. The straws are arranged in
wedges consisting of three layers rotated with respect to each other to give three-dimensional reconstruction. Each of the two ``supermodules'' consists of four layers of such wedges. 

\subsubsection{Luminosity monitor}
\label{sec-lumi}

The measurement of luminosity at HERA II must cope with the greatly
increased synchrotron-radiation background and the higher probability for multiple bremsstrahlung photons in one beam crossing. To compensate for this,  two devices, with very different systematic uncertainties, have been constructed. Both devices use the information from a small calorimeter placed around 6 m from the interaction point which  detects the radiating electron. 

The photon calorimeter is a lead-scintillator sandwich with
a position detector consisting of strips of scintillator. 
In order to cope with the synchrotron radiation background, 
an ``active filter'', consisting of
two carbon absorbers, each of two radiation lengths, alternating with Aerogel Cerenkov detectors has been constructed. The absorbers protect the calorimeter
from radiation damage, while the Cerenkov detectors detect 
high-energy photons that convert in the absorbers, allowing
the calorimeter energy to be corrected and good resolution to be
recovered. 

The pair spectrometer is situated downstream of an exit window
corresponding to around 12\% of a radiation length. The
electron-positron pairs that convert therein are separated by a dipole magnet
and detected in a pair of tungsten-scintillator sandwich calorimeters. 

The ``6 m tagger'' consists of a $10 \times 10 \times 5$ cm tungsten-scintillating fibre calorimeter next to the beam-pipe and situated inside
one of the HERA magnets. 

Each of these devices uses a newly developed common electronic readout system. With the exception of the tagger, which will be installed in January 2002, all these devices have been installed and been readout. The calorimeter is
reporting luminosity values online to the accelerator physicists while the
spectrometer is currently being commissioned. It is hoped that the
reduction of systematic error that can be obtained from independent
luminosity measurements using very different techniques will allow
a precision of around 1\% to be attained. 

\subsection{Polarisation}
\label{sec-pol}

Polarisations of around 65\% have been
achieved at HERA I. It is hoped to increase the
accuracy with which the polarisation can be measured to $\delta P/P \sim$ 2\%
per bunch per minute. 
This will be achieved by a collaboration between H1, HERMES,
ZEUS and the HERA machine in the POL2000 project. The collaboration
has constructed two instruments, one to measure
the longitudinal polarisation and the other to measure the transverse polarisation. Both detect
asymmetries in back-scattered light from high-intensity polarised lasers. 

\begin{figure}[h]
\begin{center}
\epsfig{file=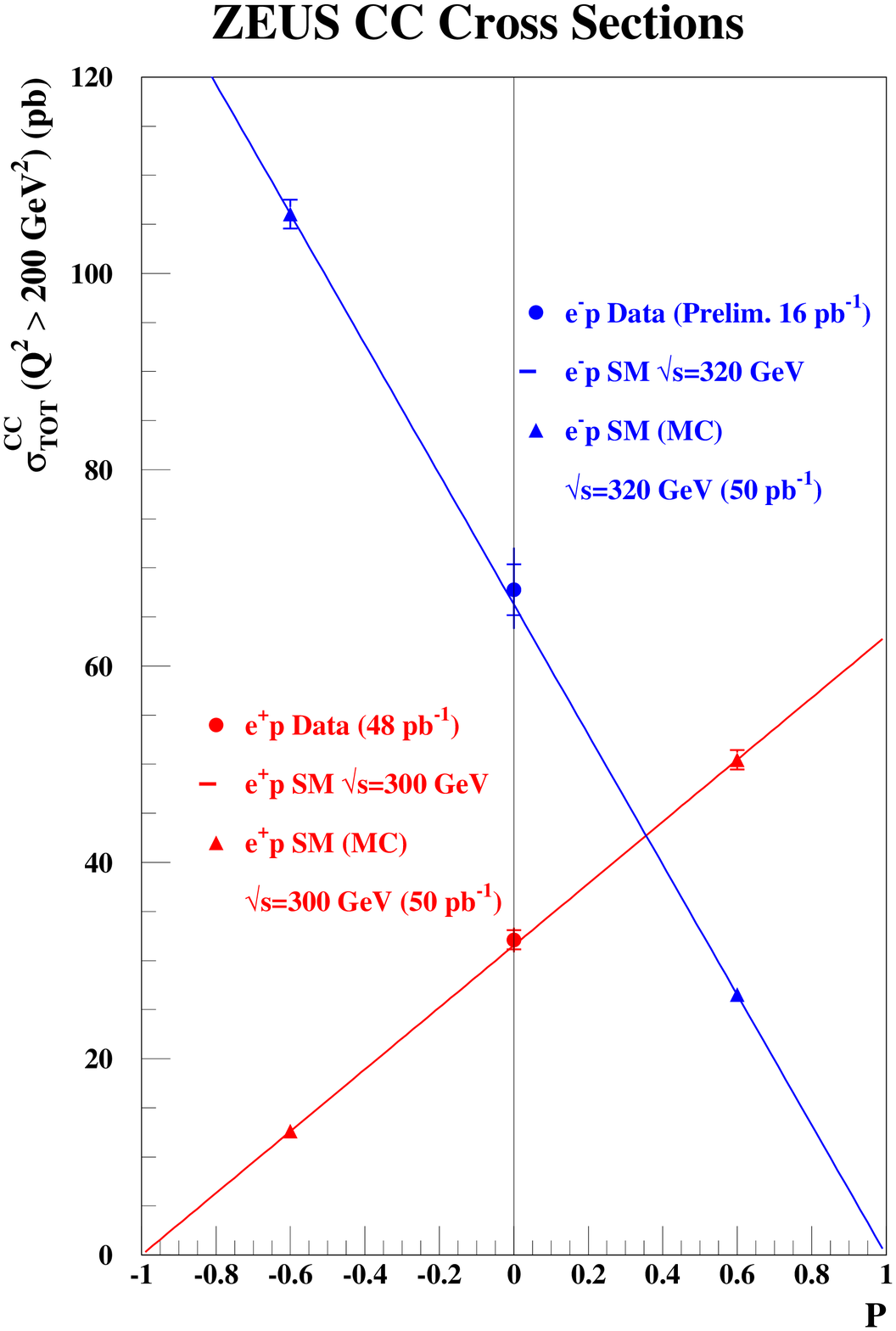,%
      width=8cm,%
      height=10cm%
        }
\end{center}
\caption{ The cross section for charged current interactions. The points at
P=0 are obtained from ZEUS preliminary results at the indicated centre-of-mass
energies, while those at non-zero polarisation are Monte Carlo simulations of 
the expected accuracy in ZEUS assuming the Standard Model cross section for an integrated luminosity of 50 pb$^{-1}$ per point.}
\label{fig:e+-pol}
\end{figure}

The combination of high-precision measurements of both luminosity and
polarisation will be important in a wide range of HERA II physics,
particularly in the electroweak sector.
The charged current cross section should vanish for the appropriate
combinations of lepton charge and polarisation. A measurement at three
polarisations, such as shown in figure~\ref{fig:e+-pol}, even with
an integrated luminosity of only 50 pb$^{-1}$ per point, will provide
an accurate test of this prediction and thereby give sensitivity
to possible new currents outside the Standard Model.

Strong polarisation effects are also predicted at high $Q^2$ in
neutral current interactions, where, e.g. at $Q^2 = 10^4$ GeV$^2$ and 
$x = 0.2$, there is a factor of two difference between the 
predicted cross sections for left- and right-handed electrons.

In addition to the use of precise luminosity and polarisation information
in the study of electroweak processes, polarisation also offers an invaluable
tool in the study of possible signals beyond the Standard Model. Varying the polarisation to reduce the
cross sections of Standard Model processes can improve the
signal to background for new physics signals, such as leptoquarks or supersymmetric particles that violate $R$ parity, for which
HERA will be competitive with the
Tevatron for the next few years. 

\section{The relevance of HERA physics to LHC experimentation}
\label{sec:HERA+LHC}

At an LHC Symposium, it is relevant to comment on the importance
of HERA physics to experimentation at LHC. 
The HERA I programme has added enormously to our understanding of the
strong interaction, in both the ``hard'' and ``soft'' regimes, and
HERA II promises similar progress in the space-like electroweak
interaction and in the search for physics beyond the Standard Model.
The distributions of partons inside the proton which are being
accurately measured at HERA are precisely those that collide at 
LHC. In addition to these obvious generic links between HERA and LHC physics,
there are some very specific areas where information from HERA impinges
directly and importantly on LHC physics. The two most important, and which
are therefore discussed here, are the determination of luminosity 
and the search for new physics processes at LHC.

\subsection{Luminosity determination at LHC}
\label{sec:LHClumi}

Much of the most exciting physics at LHC does not require an accurate
luminosity measurement; for example, the Higgs can be discovered without
any knowledge of the luminosity. However, any cross-section measurement,
the absolute determination of branching ratios and many precision
measurements do require such knowledge. The accuracy required is
relatively modest, around 5\%~\cite{MartinLHCref}. However, the task of
making a luminosity determination to even this accuracy is very
challenging. 

There are three main ways that have been discussed to determine the 
luminosity~\cite{LHCworkshop}:
\begin{enumerate}
\item measure the total and elastic cross sections;
\item measure QED lepton-pair production
\item measure inclusive $W$ and/or $Z$ production
\end{enumerate}
Since the second method is independent of HERA information, it is not
discussed further. 

The total and elastic cross sections are dominated by soft processes
and can be well explained by Regge Theory and the Pomeron. 
The Optical Theorem links
the elastic and total cross sections
via 
\begin{equation}
\left. \frac{d\sigma_{el}}{dt}\right|_{_{t=0}}
=\sigma^2_{tot}\; \frac{(1+\rho^2)}{16\pi},
\label{eq-optical}
\end{equation}
where $\rho$ is the ratio of the real to imaginary parts of the
forward elastic amplitude. 
If both the elastic and total rates
can be measured then the luminosity can be inferred, since the
elastic rate is proportional to $\sigma^2_{\rm tot} \cdot {\cal L}$
while the total rate is proportional to $\sigma_{\rm tot} \cdot {\cal L}$.
Unfortunately, it is very difficult to measure at $t=0$ at LHC! A
model that permits an extrapolation from the smallest accessible $t$ 
($\sim 0.01$ GeV$^2$) to
$t=0$ is required. Such models exist~\cite{Khoze}, but are crucially
dependent on detailed properties of the soft Pomeron and of diffraction
and elastic scattering in general that are under intensive investigation
at HERA. 

Another problem with the first method is that the measurement of elastic
scattering is really limited to relatively low luminosities. In order to
cope with the very high luminosities required at LHC it is necessary to
use another method, which can be cross-calibrated with the first method
at low luminosities and then take over. An attractive possibility is the
third method, measuring $W$ and $Z$ production rates. Although the rate
at the highest LHC luminosities compares favourably with that of Bhabha
scattering used at LEP, it suffers from the disadvantage that our ability 
to calculate the QCD processes which produce vector bosons is significantly
inferior to the QED or Electroweak theories used for Bhabha scattering. 
Nevertheless, NNLO QCD predictions are available with an estimated accuracy
of $\sim 1$\%, much better than the required accuracy. The accuracy
is more likely to be limited by our knowledge of the strong coupling constant, $\alpha_s$, whose value is unlikely to be known to comparable precision
before LHC operation. Further requirements
are that the densities of the partons that collide to produce the vector
bosons should be accurately known at the appropriate values of $x$. In
the LHC central detectors, this corresponds to approximately $10^{-4} \leq x
\leq 10^{-1}$, a range well within the ability of HERA to determine
with the requisite errors before LHC begins operation. It transpires
that the errors on the quark densities are currently the dominant ones.
The final ingredient
necessary to use $W$ and $Z$ production reliably as a luminosity monitor
is the ability to extrapolate with confidence from the $Q^2$ range at which
the parton distribution functions are measured at HERA to the
much higher values relevant at LHC. The adequacy of the standard DGLAP
evolution has now been established at HERA for the kinematic range relevant to
boson production in the central rapidity region at LHC. However, it is as well to remember that there will certainly be
regions of phase space accessible to careful experimentation
at LHC where deviations from DGLAP
would be expected and should be observable~\cite{StirlingHelsinki}.

\subsection{Searches for new physics at LHC}
\label{sec:LHCnewphys}

The enormous factor by which background processes exceed interesting events
is the dominant consideration in experimentation at LHC and the one
that sets the performance requirements for much of the apparatus, in
particular the trigger. It is clear, therefore, that a detailed understanding
of the backgrounds, which, in contrast to previous experiments, come from
beam-beam collisions rather than single-beam interactions, is vital.
Essentially all of the background originates from QCD parton-parton interactions and is therefore once again governed by the parton distribution functions determined largely by HERA data. 

In addition to the necessity to understand the backgrounds, some of the
processes studied at HERA can also give rise to interesting signals for
new physics. In general, many of the possible new species of particles
expected within the discovery potential of LHC would be colourless and
therefore their production might well be associated with
rapidity gaps. An important example of such a process is diffractive 
production of the Higgs. Since
from Tevatron data it can be predicted that Standard Model production of
events with large rapidity gaps is down by about two orders of magnitude
compared to normal events, a large improvement in signal to noise could
be obtained provided that the diffractive Higgs production is not
similarly suppressed. Unfortunately there are several models~\cite{Khoze}, whose
predictions for the ratio between diffractive Higgs production and
the dominant $gg \rightarrow H^0$ range between $10^{-1}$ and 
$10^{-12}$. The spread between these models is predominantly due to the extent to which the production is dominated by a ``soft'' or a
``hard'' Pomeron. The elucidation of the properties of such Pomerons
is currently a subject of intensive study at HERA. The other major
factor that effects the possibility of observing the Higgs via diffractive production is the extent to which the rapidity gaps thus produced are ``filled in'' by other processes such as minimum-bias interactions from the overlapping
protons, bremsstrahlung from the partons participating in the hard collision
and radiation from the partonic constituents of the Pomeron itself. The
well-known discrepancy between the CDF rapidity-gap data and the prediction from the rate of such events at HERA assuming simple vertex factorisation
gives a good handle with which to estimate the size of such 
effects~\cite{Kaidalov}. 

\section{Future prospects for lepton-proton physics}
\label{sec-future}

The future for lepton-proton physics is at least clear for the
period up until around 2006, during which the upgraded HERA II accelerator
has an exciting physics programme, as discussed in section~\ref{sec:HERAII}.
For the period beyond that, there are several possibilities. 

There are already intensive discussions on a programme at HERA beyond
that of HERA II, unsurprisingly labelled HERA III. The programme
could include lepton-deuteron scattering, with both polarised
and unpolarised deuterons, polarised electron-polarised proton
scattering, electron-nucleus scattering, dedicated experiments on
low $x$ physics and diffraction as well as a continuation of the HERMES
programme of polarised electron on gas target scattering. The dynamic
of this programme is driven mostly by unanswered questions in low-$x$
physics, particularly questions such as saturation, as discussed in
section~\ref{sec-structurefn}, which probably cannot be solved with
HERA I data and cannot be addressed at HERA II, since the relevant kinematic region is shadowed by the superconducting quadrupoles installed inside
H1 and ZEUS. The $eN$ option
would be particularly suited to address these questions, since there
are good theoretical grounds to believe that the density of quarks
in such collisions can be substantially increased over what
is possible in $ep$ collisions. Another incentive is the desire
to enter completely unexplored kinematic regions of spin physics. 
It seems likely that an exciting physics programme can be
constructed. However, the possibility of its implementation
is closely coupled to the TESLA project, which is clearly the
first priority of the DESY laboratory. Several aspects of the HERA III
programme, in particular the polarised-proton option, would require
substantial investment of both money and accelerator physicists,
which currently seems unlikely to be available inside the laboratory
without substantial external funds becoming available.

A rather similar set of physics objectives can be addressed with 
the Electron-Ion Collider (EIC) project at RHIC at Brookhaven. 
Polarised protons are readily available; the
parameters being discussed would produce $ep$ collisions at
$\surd s \sim 100$ GeV with a luminosity around $5 \cdot 10^{32}$
cm$^{-2}$ s$^{-1}$
and $eA$ collisions at $\surd s \sim 65$ GeV with luminosity
around $6 \cdot 10^{30}$ cm$^{-2}$ s$^{-1}$. 
The advantages of this proposal with
respect to HERA III are the easy availability of polarised protons
from currently available infrastructure, acceleration of heavy nuclei
up to gold and the higher luminosity. The disadvantages are that
it is relatively expensive to build the new electron storage ring
and that the centre-of-mess energy available is considerably
lower than at HERA. Once again, extensive discussions and workshops
are underway both in the particle physics and nuclear physics
communities to discuss this project.

In the more distant future, further exciting options open up. The
neutrino factory concept would provide enormous fluxes of neutrinos
(approximately four orders of magnitude above what can be achieved
with conventional sources) which would enable a rich programme of
fixed-target neutrino physics including for the first time the
possibility of using polarised targets.

Another possibility is to construct 
$ep$ colliders with higher energy than HERA. The THERA proposal
would utilise the electrons from the TESLA superconducting
linac to collide with protons in the HERA ring. By using
the full length of the linac, electrons of up to 500 GeV could
be collided, producing a centre-of-mass energy of 1.35 TeV.
Such a machine would extend the HERA kinematic range by 
two orders of magnitude in $Q^2$ and one in $x$. Exploration
of the $x$ range around $10^{-6}$ with $Q^2$ well into the
perturbative QCD regime would become possible, allowing a
comprehensive investigation of all the problems in low-$x$
physics, saturation and diffraction discussed above. It would
also be an ideal machine to measure parameters of any possible
new state with leptoquark-like properties, such as many
states in R-parity-violating supersymmetry. THERA also
has discovery potential for certain classes
of possible new states, for example right-handed or excited 
neutrinos.

The major limitation with THERA is the attainable luminosity,
which is limited by colliding electron bunches from the linac
rather than in a storage ring. It may be possible to get up
to $10^{31}$ cm$^{-2}$ s$^{-1}$, but it will not be easy,
and $10^{30}$ cm$^{-2}$ s$^{-1}$ would be a more conservative 
value. This will certainly be adequate for the low-$x$ physics,
but will limit studies at the highest $Q^2$. Such a 
luminosity limitation is avoided in the ``LHC $\times$ LEP'' machine, in which
new magnets could be built to reinstate the LEP ring and
the electrons collided with protons in LHC. Although this option
has been left open at CERN, at the moment it is not being actively
pursued.

\section{Summary}
\label{sec:summary}

In the last decade, HERA I has changed our perception of QCD out
of all recognition. In many cases the precision of the data 
mandate NNL, or even high order, QCD predictions. The study
of diffraction and the transition region between soft and
hard physics may be beginning the era of of quantitative study
of the central problem of the strong interaction, confinement.

The precision attained on the parton distribution functions
in the proton, mostly with HERA data, will directly influence
many of the physics topics at LHC. In addition, some of the 
processes studied at HERA may well lead to distinctive signatures
for new physics and contribute to the luminosity determination
necessary for many precision studies.

Now that the first collisions have been achieved in HERA II with
specific luminosities close to the design, the immediate future
for $ep$ physics seems bright. Ideas for further developments
such as HERA III and EIC are already gathering momentum, and
possibilities for the distant future are also in place. These
latter will surely be thrown into a much clearer light by
what we all hope and expect to be revolutionary discoveries at LHC.
In any case, $ep$ physics has had a glorious past and I am
confident will continue to produce much excitement in the future.
          
\section*{Acknowledgements}
I am grateful to Gino Saitta and his colleagues from INFN Cagliari
and Roma for organising a splendid
workshop in beautiful surroundings. I thank Carlo
Bosio for his patience with numerous deadline postponements for this
manuscript. I am grateful to Masahiro Kuze for a careful reading of
the manuscript; as usual, he improved both my physics and my grammar.

\end{document}